\documentstyle[12pt,aasms4,graphics]{article}

\def\baselinestretch{1.4}
\begin{document}
\oddsidemargin=0mm

\title{
On the Progenitor of the Type II Supernova 2004et in NGC 6946\footnote{Based on data obtained at the Canada-France-Hawaii Telescope}
}

\author{Weidong Li\altaffilmark{1}, Schuyler D. Van Dyk\altaffilmark{2}, 
Alexei V. Filippenko\altaffilmark{1}, and Jean-Charles 
Cuillandre\altaffilmark{3} \\
Email: (wli, alex)@astro.berkeley.edu, vandyk@ipac.caltech.edu, jcc@cfht.hawaii.edu}

\altaffiltext{1}{Department of Astronomy, University of California, Berkeley,
CA 94720-3411.}

\altaffiltext{2}{Spitzer Science Center, California Institute of Technology, 
Mailcode 220-6, Pasadena, CA 91125}

\altaffiltext{3}{Canada-France-Hawaii Telescope Corporation, 65-1238 Mamalahoa 
Hwy, 
Kamuela, HI 96743 }

\slugcomment{Submitted to PASP}

\begin{abstract}

Supernova (SN) 2004et is the eighth historical SN in the nearby spiral galaxy
NGC 6946.  Here we report on early photometric and spectroscopic monitoring of
this object. SN 2004et is a Type II event, exhibiting a plateau in
its light curves, but its spectral and color evolution appear to differ
significantly from those of other, more normal Type II-plateau (II-P) SNe.  We
have analyzed Canada-France-Hawaii Telescope (CFHT) images of the host
galaxy taken prior to the SN explosion, identifying a candidate progenitor for
the SN.  The star's absolute magnitude and intrinsic color imply that it was a
yellow, rather than red, supergiant star, with an estimated zero-age main
sequence mass of $15^{+5}_{-2}\ M_\odot$.  Although this mass estimate is
consistent with estimates and upper limits for the progenitors of other, more
normal SNe~II-P, the SN 2004et progenitor's unusual color could further imply a
pre-explosion evolutionary history analogous to, but less extreme than, that
for the progenitors of the peculiar Type II-P SN 1987A or the Type IIb SN
1993J.  The identity of the progenitor candidate needs to be verified when the
SN has significantly dimmed.

\end{abstract}


\section{Introduction}

   Identification of the progenitors of supernovae (SNe) provides direct
information on their explosion mechanisms, a key issue in studies of SNe. The
white dwarfs thought to give rise to SNe~Ia have such exceedingly low
luminosities that they cannot be detected in other galaxies, and a direct
identification of a possible binary companion has yet to be made (but see
Ruiz-Lapuente et al. 2004). Core-collapse SNe, on the other hand, come from
more luminous, massive stars. Unfortunately, even these progenitors are so
faint that detection (ground-based or space-based) is confined to the most
nearby galaxies, in which SN discoveries are relatively rare.  Up to now, only
half a dozen SNe have had their progenitors identified: SN 1961V in NGC 1058
(Zwicky 1964, 1965), SN 1978K in NGC 1313 (Ryder et al. 1993), SN 1987A in the
Large Magellanic Cloud (LMC; e.g., Gilmozzi et al. 1987; Sonneborn, Altner, \&
Kirshner 1987), SN 1993J in NGC 3031 (M81; Aldering, Humphreys, \& Richmond
1994; Cohen, Darling, \& Porter 1995), SN 1997bs in NGC 3627 (M66; Van Dyk et
al. 1999, 2000), and SN 2003gd in NGC 628 (M74; Van Dyk, Li, \& Filippenko
2003c; Smartt et al. 2004).  It should be noted that all of these core-collapse
SNe were somewhat unusual, except for the normal Type II SN 2003gd.

   Additionally, the normal Type II-P SN 2004dj in NGC 2403 was found to occur at
a position coincident with a compact star cluster (Ma\'{\i}z-Apell\'aniz et
al. 2004). By studying the stellar population and the age of the cluster, an
estimate of a main-sequence mass of 15 $M_\odot$ for the SN progenitor was
made. Tentative identification and upper mass limits are also derived for
several other SNe (Van Dyk, Li, \& Filippenko 2003a, 2003c; Leonard et
al. 2002a, 2003; Smartt et al. 2001, 2002).

   Here we attempt to identify the progenitor of the Type II SN 2004et in NGC
6946.  The SN was discovered by Moretti (2004) at about 12.8 mag on unfiltered
CCD images taken with a 0.4-m telescope on Sep. 27 (UT dates are used
throughout this paper), with a reported position of $\alpha$(J2000) =
$20^h35^m25{\fs}33$, $\delta$(J2000) = $+60^\circ07\arcmin17{\farcs}7$. A
high-resolution optical spectrum taken by Zwitter \& Munari (2004) suggested
that SN 2004et is a Type II event, which was subsequently confirmed by a
low-resolution optical spectrum (Filippenko et al. 2004).  We note that the
host galaxy, NGC 6946 (Arp 29), a nearly face-on ($i \approx 29{\fdg}5$),
starbursting spiral galaxy at relatively low Galactic latitude ($b^{\rm II}
\sim 11{\fdg}5$), is an especially prodigious SN producer: SN 2004et is the
eighth historical SN (including SNe 1980K and 2002hh) in this galaxy.

   The explosion date of SN 2004et is well constrained. Yamaoka \& Itagaki
(2004) reported that nothing was detected to a limiting magnitude of 18.5 at
the position of SN 2004et on Sep. 19.655. Klotz, Pollas, \& Boer (2004)
reported that NGC 6946 was imaged frequently by the robotic TAROT telescope.
The SN was not detected on Sep. 22.017 (limiting mag 19.4$\pm$1.2), but was
detected at mag 15.17$\pm$0.16 on Sep. 22.983, only about one day later. The SN
then brightened to mag 12.7$\pm$0.3 mag on Sep. 25.978. We will use Sep. 22.0
(JD 2453270.5) as the time of explosion for SN 2004et throughout this paper.

   A preliminary study of the progenitor of SN 2004et, based on images taken
with the 0.9-m Kitt Peak telescope (see Van Dyk, Hamuy, \& Filippenko 1996),
was reported by Li, Filippenko, \& Van Dyk (2004). These images were taken
under relatively poor conditions (seeing about $2{\farcs}7$), and with low
resolution ($0{\farcs}863$ pixel$^{-1}$). A faint object, considered possibly
extended, was detected at the position of SN 2004et in the $R$-band image,
though the high measured luminosity of the object led to the suggestion that it
might be a star cluster, rather than a single star. Analysis of deeper,
higher-resolution (binned to $0{\farcs}412$ pixel$^{-1}$) images taken under
better seeing conditions ($\sim 0{\farcs}8$) with the Canada-France-Hawaii
Telescope (CFHT) was also reported by Li et al. (2004).  Their revised
luminosity of the progenitor was consistent with a massive supergiant, although
likely too bright and too blue for a single red supergiant.

   In this paper we analyze the best available CFHT images of the site of SN
2004et in detail, and the results here supersede those reported by Li et
al. (2004).  Section 2 describes spectroscopic and photometric observations of
SN 2004et itself.  Section 3 describes our analysis of these pre-SN CFHT
images.  Discussion of the likely nature of the SN progenitor, based on this
analysis, is in Section 4, and our conclusions are summarized in Section 5.

\section{Observations of SN 2004et}

   From the early photometric and spectroscopic observations of SN 2004et we
can get at least an initial indication of its nature.

   SN 2004et has been monitored in the Johnson-Cousins $UBVRI$ system with the
0.76-m Katzman Automatic Imaging Telescope (KAIT; see Li et al. 2000;
Filippenko et al. 2001) at Lick Observatory since its discovery.  The images
were reduced using standard aperture photometry in the IRAF\footnote{IRAF
(Image Reduction and Analysis Facility) is distributed by the National Optical
Astronomy Observatories, which are operated by the Association of Universities
for Research in Astronomy, Inc., under cooperative agreement with the National
Science Foundation.} DAOPHOT package (Stetson 1987), and transformation to the
standard Johnson-Cousins $UBVRI$ system was performed with the local comparison
stars listed in Table 1; see Li et al. (2001) for more details.  Figure 1 shows
an $I$-band image of the SN, with these comparison stars labeled.  The final
photometry for SN 2004et is listed in Table 2, while Figure 2 shows the light
curves of SN 2004et (solid circles), together with comparisons to the typical
Type II-P SN 1999em (lines; Leonard et al. 2002b).  The pre-discovery $R$-band
magnitudes reported by Klotz et al. (2004) are also shown.  The light curves of
SN 1999em have been shifted arbitrarily by eye to match those of SN 2004et. The
two SNe have rather similar light curves in the $V$, $R$, and $I$ bands,
suggesting that SN 2004et is a relatively normal SN~II-P.  However, SN 2004et
appears to evolve more slowly than SN 1999em in the $U$ and $B$ bands.

   The color curves of SN 2004et are shown in Figure 3 (symbols), together with
comparisons to SN 1999em (lines; Leonard et al. 2002b).  The colors of SN
1999em were dereddened by the estimated extinction of $E(B - V)$ = 0.10 mag,
following the reddening law in Schlegel, Finkbeiner, \& Davis (1998). The color
curves of SN 2004et were shifted by eye to match those of SN 1999em, and the
best fit was found when the curves were shifted 0.0 mag, $-$0.46 mag, $-$0.35
mag, and $-$0.28 mag for the $U-B$, $B-V$, $V-R$, and $V-I$ colors,
respectively.  If we assume SNe 2004et and 1999em have the same color
evolution, the offsets in these color curves suggest an $E(B-V)$ color excess
for SN 2004et of 0.0 mag, 0.46 mag, 0.55 mag, and 0.20 mag, respectively. The
$B-V$ curve yields a color excess that is consistent with the value derived
from high-resolution spectroscopy [$E(B - V)$ = 0.41 mag; see \S 3].  The color
evolution of SNe 2004et and 1999em clearly differ from each other.  In
particular, the $U-B$ and $B-V$ colors of SN 2004et evolve more slowly than
those of SN 1999em, in agreement with the trend seen in the $UB$ light curves.

   We have obtained two optical spectra of SN 2004et using the Lick Observatory
3-m Shane telescope with the Kast double spectrograph (Miller \& Stone 1993) on
2004 Oct. 1 and Oct. 12, 9 and 20 days after the SN explosion,
respectively. The journal of observations is listed in Table 3. The data were
reduced using standard techniques as described by Li et al. (2001) and
references therein.  Flatfields for the red CCD were taken at the position of
the object to reduce near-IR fringing effects.  The spectra were corrected for
atmospheric extinction and telluric bands (Bessell 1999; Matheson et al. 2000),
and then flux calibrated using standard stars observed at similar airmass on
the same night as the SN.

    Figure 4 shows the spectra of SN 2004et, with a comparison to SN 1999em at
similar epochs (Hamuy et al. 2001). All spectra of SN~1999em have been
dereddened by $E(B - V) = 0.10$ mag (Leonard et al. 2002b), and all spectra of
SN 2004et have been dereddened by $E(B - V) = 0.41$ mag (see \S 3).  The
spectra have also been corrected for the redshift of the SN host galaxy ($z =
0.000160$ for SN~2004et, and $z = 0.002392$ for SN~1999em). The spectra of SN
2004et show some clear differences and peculiarities relative to SN 1999em. The
spectrum at 9 days after explosion has an overall bluer continuum than that of
SN 1999em in the range 4000--10000~\AA, and there is a peculiar decline
blueward of 4000~\AA\ not commonly observed in the spectra of normal SNe II-P.
The P-Cygni-like profile of H$\alpha$ is dominated by the emission component,
although the other H Balmer lines have more typical P-Cygni profiles.  An
additional absorption feature appears to be present just blueward of the
He~I/Na~I~D absorption at 5700~\AA.

   The SN 2004et spectrum at 20 days after explosion is also bluer than that of
SN 1999em.  A shallow absorption trough, which appears to consist of two
components, developed at the absorption component of the P-Cygni profile of
H$\alpha$, but the emission component is still predominant.  In comparison, SN
1999em has a more typical P-Cygni H$\alpha$ profile. While strong Fe~II lines
appear at the blue end of the SN 1999em spectrum, these lines are less
prominent in SN 2004et. Overall, SN 2004et seems to evolve more slowly than SN
1999em, especially in the UV part of the spectrum.  This is consistent with the
slower $UB$ photometric evolution for SN 2004et, discussed above.

   Clearly, SN 2004et is of Type II and (presumably) a core-collapse SN.
However, we conclude that although SN 2004et shows a plateau in its light curve
(Figure 2), it displays noticeable peculiarities in both its spectroscopic and
photometric properties, when compared to the canonical SN~II-P. The relative 
blue colors and weak P-Cygni absorption features of SN 2004et may indicate 
the presence of a circumstellar interaction, as discussed more in \S 4.
More data for
the SN from continuing follow-up observations will provide additional
information on the nature and evolution of SN 2004et.

\section{Pre-Supernova Images}

   Images of NGC 6946, containing the SN 2004et site, were obtained by one 
of the authors (JCC) with the
CFHT on two occasions prior to the explosion: (a) broad-band $B$, $V$, and $R$
images were obtained with the CFH12K mosaic camera on 2002 Aug. 6, and (b)
Sloan $u\arcmin$, $g\arcmin$, and $r\arcmin$ images were obtained with the
MegaCam camera on 2003 Oct. 23.  A summary of these pre-SN CFHT data is given
in Table 4.  The 2$\times$2 binned images of dataset (a) were those analyzed
and reported by Li et al. (2004). Here, we analyze the full-resolution
(unbinned) images.  Both the CFH12K and the MegaCam cameras have large fields
of view.  To facilitate the analysis, here only the image sections that contain
the SN site are studied.  For dataset (a), a section of $6{\farcm}8 \times
6{\farcm}8$ around SN 2004et is considered, while for dataset (b), the section
is $2{\farcm}5 \times 2{\farcm}5$ around the SN site.

   It is essential to locate with high astrometric precision the SN site in the
CFHT images.  We have adopted two methods for this purpose: an astrometric
solution and image-to-image registration.  For our first approach we use
several images of SN 2004et obtained with KAIT and derived astrometric
solutions, based on the accurate positions for stars in the USNO-A2.0 catalog
seen in the SN 2004et field. We have measured an accurate position for the SN
of $\alpha$(J2000) = $20^h35^m25{\fs}37$, $\delta$(J2000) =
$+60^\circ07\arcmin17{\farcs}8$, with a total uncertainty of about
$0{\farcs}2$. This is in excellent agreement (differing by $0{\farcs}07$ in
$\alpha$, and $0{\farcs}1$ in $\delta$) with the position measured from the
radio observations (Stockdale et al. 2004).

   We apply exactly the same procedure to the pre-SN CFHT images and inspect
the immediate site of SN 2004et. We identify an apparent stellar object at the
position of SN 2004et, especially in the $R$, $g\arcmin$, and $r\arcmin$
images. The position of this object, measured from the best-seeing $r\arcmin$
image, is $\alpha$(J2000) = $20^h35^m25{\fs}38$, $\delta$(J2000) =
$+60^\circ07\arcmin18{\farcs}0$, again with a total uncertainty of about
$0{\farcs}2$. This position is very consistent (differing by $0{\farcs}07$ in
$\alpha$ and $0\farcs2$ in $\delta$) with that measured for the SN from the
KAIT images, to within the uncertainties.

   For the image registration approach, we choose the best available KAIT
image, detect all the stars with sufficient signal-to-noise ratio, and output
their $(x,y)$ positions to a list.  The same is done for the CFHT images. The
two sets of star lists are then used to solve for the geometrical
transformation between the two sets of images.  In essence, both of our
approaches are quite similar, the difference being that image registration
finds the geometrical match between images directly, whereas the astrometric
solution uses an external reference frame (the USNO-A2.0 catalog). We therefore
might expect the image registration method to perform somewhat better, since
two images are compared directly.

   Figure 5 shows the results from the image registration method, when
the CFHT $R$-band image is compared to the KAIT $I$-band image shown in Figure
1.  The geometrical solution has a root-mean-square (rms) of about
$0{\farcs}1$.  When the position of the SN measured from the registered KAIT
image is mapped onto the CFHT image, it is offset from the object we identify
by only $0{\farcs}08$.  As shown in the right panel of Figure 5, the progenitor
star is right in the middle of the 2$\arcsec$ radius circle centered on the SN
position. (Such a large circle is shown only to guide the eye; we measured the
SN position with a much lower uncertainty of about $0{\farcs}2$.)

   At the distance of NGC 6946, the uncertainty in the SN position
($0{\farcs}2$) corresponds linearly to $\sim$5 pc, a radial scale in a normal,
star-forming spiral galaxy within which many massive stars could exist.
However, we note that no other luminous stellar sources are detected within the
$0{\farcs}2$ error circle of the SN position. The possibilities, then, are that
either the star we identify here is the progenitor of SN 2004et, or the
progenitor is not detected in the CFHT images. It is also possible that the
object we have identified is actually a compact star cluster. More discussion
of these various possibilities is given in \S 4.  However, at least for the
sake of argument, hereafter we assume that the identified star is the
progenitor of SN 2004et. We further investigate its luminosity and color from
the photometry, to determine its nature.

   Figure 6 shows the SN 2004et environment in all of the CFHT images. The
progenitor is well detected in the $R$, $g\arcmin$, and $r\arcmin$ images. It
is also detected on the $V$ image, but its profile is not well defined.  The
progenitor is marginally detected in the $B$ image and may be only very
tentatively detected in the $u\arcmin$ image.

  We derive instrumental magnitudes for the progenitor star via
point-spread-function (PSF) fitting photometry in IRAF/DAOPHOT.  A PSF that
varies linearly across the image is constructed for each frame using all
available bright and isolated stars. Because of the relatively low Galactic
latitude ($b^{\rm II} \sim 11{\fdg}5$) of NGC 6946, there are plenty of
stars to sample the PSF across the field, and usually more than 20 stars
are used. To reduce the
contamination from neighboring stars, the PSF is first fit and subtracted from
the two stars seen adjacent to the progenitor in Figure 6. To avoid any
possible random positional shifts in the PSF-fitting procedure, the
progenitor's centroid is fixed, using relative offsets from a nearby bright
star (marked by a square in the right panel of Figure 5)
in the $R$ and $r\arcmin$ images, in
which the progenitor is well detected. The model PSF is then fit at this
position, and the adopted instrumental magnitudes for the star result.

   Since the progenitor is faint and is located in a rather complex region, the
instrumental magnitudes resulting from IRAF/DAOPHOT may have rather large
uncertainties. To establish a more realistic assessment of the true photometric
uncertainty, we have applied a procedure that is similar to the Monte Carlo
simulation performed by Riess et al. (1998): we first introduce artificial
stars with the same brightness as the progenitor throughout the image, and then
photometrically recover these artificial stars using the PSF-fitting procedure.
For each grid in the 21$\times$21 pixel rectangle region centered on the
progenitor, we superimpose an artificial star, for a total of 441
measurements. The photometric uncertainty is then calculated from the standard
deviation for the ensemble of 441 measurements, the average of the photometric
uncertainties from IRAF/DAOPHOT, and the standard deviation of the
uncertainties output from IRAF/DAOPHOT, all added in quadrature.

   Photometric calibrations of the NGC 6946 field in the $B$, $V$, $R$, and $I$
filters were obtained on 2002 Nov. 5, 2003 May 31, and 2003 June 1 with the
Nickel 1-m reflector at Lick Observatory. Additional calibrations, including
$U$, were obtained on 2004 Oct. 12 and 14 with KAIT. However, the field has not
been directly calibrated in $u\arcmin$, $g\arcmin$, and $r\arcmin$.  Since the
transformation between the Johnson-Cousins $UBVRI$ and the Sloan $u\arcmin
g\arcmin r\arcmin i\arcmin z\arcmin$ photometric systems has been well studied
(e.g., Fukugita et al. 1996; Krisciunas, Margon, \& Szkody 1998), we adopt the
transformation from Krisciunas et al. (1998) to obtain the $u\arcmin g\arcmin
r\arcmin$ calibration for those CFHT images. The photometry for the local
comparison stars in the Sloan bands is listed in Table 1.  We have verified the
transformation between the two photometric systems, based on the stars in Table
1, and found the transformation to be satisfactory to $\pm$0.03 mag.

   The instrumental magnitudes of the progenitor are then transformed to the
standard Johnson-Cousins and Sloan systems by performing relative photometry
between the progenitor and the comparison stars in Table 1.  Stars 1 through 6
are used for all the images, except $u\arcmin$, in which only Stars 7 and 8 are
sufficiently bright in the KAIT $U$ images to provide a reliable
transformation. To compare the photometry from the two datasets, we transformed
the $u\arcmin g\arcmin r\arcmin$ magnitudes of the progenitor to $UBVR$ and
calculated the differences between the two sets of photometry. The apparent
magnitudes for the progenitor in $UBVR$ are given in Table 5.  The
uncertainties in these magnitudes are based on the photometric uncertainty of
the instrumental magnitudes, described above, and the standard deviation in the
photometric calibration using the comparison stars, added in quadrature. The
standard deviation in the transformation is also included for the $UBVR$
magnitudes converted from the $u\arcmin g\arcmin r\arcmin$ magnitudes.

   Analysis of Table 5 suggests that the $BVR$ magnitudes measured from the
2002 image set are consistently brighter than those transformed from the
$u\arcmin g\arcmin r\arcmin$ magnitudes measured from the 2003 image set. The
differences are about 0.9$\sigma$, 2.1$\sigma$, and 1.6$\sigma$ for $B$, $V$,
and $R$, respectively.  While the differences are within the 3$\sigma$
uncertainties, the fact that all measurements from one dataset are
systematically brighter than the other is somewhat disconcerting.  Several
factors may contribute to these differences: (1) the seeing and resolution are
better for the $u\arcmin g\arcmin r\arcmin$ images, which could result in less
background contamination and, therefore, overall fainter measurements; (2) the
progenitor may be yellow in color (see below) and is relatively bright in $V$,
and, since there is not a good matching filter in the Sloan system for $V$, the
transformation from the $g\arcmin r\arcmin$ magnitudes to $V$ may be
unreliable; and (3) the progenitor does not have a well-defined stellar profile
in the $V$ image (Figure 6), and the uncertainty in the $V$ magnitude is
underestimated, even with our extensive Monte Carlo experiments.

\section{The Progenitor of SN 2004et}

   We can estimate the mass of the SN progenitor by comparing the intrinsic
color and absolute magnitude of the object with stellar evolution tracks of
massive stars having different zero-age main-sequence masses ($M_{\rm ZAMS}$).
In \S 3 we discussed the apparent photometry for the progenitor (see Table 5).
To estimate the absolute magnitude of the progenitor, we need to know the
distance to NGC 6946 and an estimate of the extinction toward SN 2004et. The
distance to NGC 6946 is measured to be 5.5$^{+1.1}_{-0.9}$ Mpc from the H~I
Tully-Fisher relation (Pierce 1994), 5.4 Mpc from the CO Tully-Fisher relation
(Schoniger \& Sofue 1994), and 5.7$^{+0.7}_{-0.7}$ Mpc from the Expanding
Photosphere Method for Type II SNe (Schmidt et al. 1994). Here we adopt
5.5$\pm$1.0 Mpc for the distance. The relatively large uncertainty in the
distance alone results in an uncertainty of $\pm$0.4 mag for the absolute
brightness (luminosity) of stars in NGC 6946.

   In \S 2 we showed that SN 2004et and the typical SN II-P 1999em are unlikely
to have the same color at a given age, and therefore a reliable extinction
estimate for SN 2004et cannot be obtained from a comparison of the color curves
for these two SNe.  However, the spectrum obtained by Zwitter \& Munari (2004)
shows distinct interstellar Na~I absorption lines at heliocentric radial
velocities of $-21.1 \pm 0.7$ and $+45.2 \pm 0.4$ km s$^{-1}$, matching the
expectation for a separate origin, respectively, within the Milky Way Galaxy
and NGC 6946 (the systemic velocity of this galaxy is 48 km s$^{-1}$;
NED\footnote{NED is the NASA/IPAC Extragalactic Database,
http://nedwww.ipac.caltech.edu.}).  The measured equivalent width of the Na~I~D
lines corresponds to a total reddening of $E(B - V)$ = 0.41 mag, following the
calibration by Munari \& Zwitter (1997).  The Galactic component for the
reddening alone is estimated to be $E(B - V)$ = 0.34 mag (Schlegel et
al. 1998). We adopt $E(B - V)$ = 0.41$\pm$0.07 mag for the reddening toward SN
2004et, the lower limit of which corresponds to no host-galaxy reddening to SN
2004et within NGC 6946.

   Correcting for the distance and extinction to SN 2004et, we derive the
absolute magnitude and intrinsic color of the progenitor given in Table 7. To
the best of our knowledge no appropriate stellar evolution tracks in the Sloan
$u\arcmin g\arcmin r\arcmin i\arcmin z\arcmin$ system exist, so we are forced
to consider only tracks in the Johnson-Cousins system.  As a result of large
uncertainties in the distance estimate and the photometry of the progenitor,
the absolute magnitudes have quite large uncertainties as well (0.48 to 0.78
mag).

   The color-magnitude diagrams (CMDs) of massive stars are significantly
affected by the adopted metallicity, so we attempt to constrain the metallicity
of the SN 2004et environment, as measured by other investigators.  SN 2004et
occurred at about $273\arcsec$ southeast of the nucleus of NGC 6946, or at
$R/R_{25} \approx 0.8$. Zaritsky, Kennicutt, \& Huchra (1994) published the
metallicity and its radial gradient in NGC 6946 using the $R_{23}$ method
(Pagel et al. 1979).  At the radial distance of SN 2004et, the relative oxygen
abundance log(O/H) + 12 would be 8.84 dex, very close to the solar value (8.8
dex; Grevesse \& Sauval 1998). However, Pilyugin et al. (2002) also determined
the metallicity in NGC 6946 using the $P$-method (Pilyugin 2000). At the radial
distance of SN 2004et, the relative oxygen abundance from the Pilyugin et
al. measurement is 8.34 dex, which is only one-third solar.  Since we consider
both metallicity estimates to be equally valid, we use both of them in our
analysis.

   Figure 7 shows the CMDs for SN 2004et, compared with model stellar evolution
tracks for a range of masses from Lejeune \& Schaerer (2001), assuming enhanced
mass loss for the most massive stars and solar metallicity ($Z = 0.02$). Figure
8 illustrates the results for a metallicity $Z = 0.008$ (40\% of solar).  The
open square shows the locus of the progenitor in the diagrams derived from the
$BVR$ images, while the solid square gives the locus from the $u\arcmin
g\arcmin r\arcmin$ images.

   From both Figures 7 and 8 we can rule out that the progenitor was a blue
star (with the possible exception for the [$M^0_V$, $(V-R)_0$] and [$M^0_R$,
$(V-R)_0$] CMDs measured from 2002 CFHT dataset, where the $(V-R)_0$ color has
a significantly larger uncertainty than that from 2003 dataset).  Both image
datasets, however, suggest that the progenitor may, in fact, be a yellow
supergiant (YSG). Generally, we expect the progenitors of normal SNe~II-P to be
red supergiants (RSGs), since the optical P-Cygni spectral line profiles and
especially the plateau phase of the light curve (arising from a hydrogen
recombination wave in the envelope) require such extensive hydrogen envelopes.
As discussed in \S 2, however, SN 2004et does not have the typical P-Cygni
profiles, especially for the strong H$\alpha$ line. 

   The estimates for the mass of the progenitor ($M_{\rm ZAMS}$) derived by eye
from the CMDs are listed in Table 7. The upper limit with entry ``$?$'' means
that the value is uncertain.  The median value for each metallicity is also
listed. For $Z = 0.02$, we note that the mass estimated from the 2002 image
dataset (median = $16^{+6}_{-4}\ M_\odot$) is systematically larger than that
from the 2003 dataset (median = $13^{+4}_{-2}\ M_\odot$), though the estimates
overlap within the errors.  When subsolar metallicity ($Z = 0.008$) is assumed,
the mass estimate from the 2002 dataset decreases by $1 M_\odot$ (median =
$15^{+6}_{-2}\ M_\odot$), while the mass estimate from the 2003 dataset
increases by $1 M_\odot$ (median = $14^{+2}_{-2}\ M_\odot$), resulting in the
two estimates being more consistent with each other.  When all mass estimates
are considered, the median value is $M_{\rm ZAMS} = 15^{+5}_{-2}\ M_\odot$.

   This mass estimate for the progenitor argues against the detected object
being a compact star cluster, whose mass would be several orders of magnitude
larger (Ma\'{\i}z-Apell\'aniz et al. 2004).  We cannot rule out, however, that
the object is composed of several stars, or is a binary system. One possible
scenario is that the object is a RSG and a blue supergiant (BSG) pair, possibly
interacting (as may have been the case for the SN 1993J progenitor; e.g., Van
Dyk et al. 2002; Maund et al. 2004), whose combined color is yellow.  Our
absolute magnitude estimates could be as high as $\sim$0.5 mag brighter than a
single supergiant star.  In fact, if $M^0_V \approx -7$ mag and the initial
mass range is $\sim$13--20 $M_\odot$ for the SN 2004et progenitor, this is
consistent with what Van Dyk et al. (2002) found for the Type IIb SN 1993J,
although no indications exist that SN 2004et is of Type IIb.

   Additionally, as seen in both Figures 7 and 8, some massive single stars are
expected to evolve between the BSG and RSG phase more than once during their
lives, especially at lower metallicities.  Theoretically, it is possible that
the progenitor evolved off the main sequence, to the BSG phase, and to a YSG
phase, before explosion.  Alternatively, after evolving from the BSG to a RSG
phase, it exploded as a YSG on its way back to the BSG phase.  Finally, the
star may have evolved from being a BSG to a RSG, then back to the BSG phase,
then to a YSG, when it exploded.  Such evolutionary ``loops'' from BSG to RSG
to BSG evolution are what have been invoked to explain the blue progenitor of
SN 1987A in the LMC (see Arnett et al. 1989; however, see also Podsiadlowski et
al. 1993).  

   One important discriminator among these various evolutionary scenarios is
the amount of circumstellar material (CSM) around the progenitor, expelled as a
stellar wind, at the time of explosion.  If the progenitor spent any
appreciable time as a RSG before explosion, dense CSM might be expected.  Radio
emission detected from SN 2004et on Oct. 5 (Stockdale et al. 2004), just 14
days after explosion, suggests the presence of appreciable CSM around SN
2004et, so it is quite possible that the progenitor had experienced a RSG stage
close to the time of explosion.  Further modelling of both the optical and
radio properties of SN 2004et is clearly required.

   It is also possible that the SN 2004et progenitor is indeed a RSG, correctly
identified in the $R$ and $r\arcmin$ images, but that the star we identified in
the bluer bands is another, neighboring object (or the RSG is heavily
contaminated by other sources).  It is also possible that the SN 2004et
progenitor has not been detected in the CFHT images, and the object we have
identified has no direct connection at all to the SN.  Unfortunately, the
quality of the CFHT images, although high, is not sufficient to eliminate
either of these possibilities. High spatial resolution images of the SN 2004et
environment, if obtained with the {\sl Hubble Space Telescope} or with
ground-based adaptive optics systems when the SN has
significantly dimmed, could provide us with more definitive answers regarding
the various possibilities.  In addition, high spatial and spectral resolution
observations of a sufficiently dimmed SN might reveal the presence of a binary
companion, as appears to be the case for SN 1993J (Maund et al. 2004).

\section{Conclusions}

   Spectroscopic and photometric observations of SN 2004et in NGC 6946 show
that it is of Type II. Although the SN exhibits a plateau phase in its light
curves, noticeable differences exist in its spectral and color evolution, when
compared to more normal SNe~II-P.

   By analyzing the CFHT images of the SN site taken before explosion, we have
identified a candidate SN progenitor.  The progenitor star appears to be a
yellow supergiant with an estimated ZAMS mass of $15^{+5}_{-2}\ M_\odot$, and
it may have experienced a red supergiant stage prior to explosion.  It is also
possible that the progenitor was an interacting binary system, consisting of a
red and blue supergiant, similar to the SN 1993J progenitor (Van Dyk et
al. 2002).  If this star is indeed the SN progenitor, it is only the seventh
such progenitor ever directly identified.  The mass estimate for the SN 2004et
progenitor is consistent with the limits on the progenitor masses for three
other SNe~II-P: SN 1999gi ($\lesssim 15^{+5}_{-3}\ M_\odot$; Leonard et
al. 2002a), SN 1999em ($\lesssim 20\pm5$ M$_\odot$; Leonard et al. 2003), and
SN 2001du ($\lesssim 13^{+7}_{-4}$ M$_\odot$; Van Dyk et al. 2003c). However,
it is somewhat higher than the derived progenitor mass for SN 2003gd ($\sim
8$--9 $M_\odot$; Van Dyk et al. 2003c; Smartt et al. 2004).

\acknowledgments

   We thank Matt Malkan and Tommaso Treu (University of California, Los
Angeles) for obtaining optical spectra of SN 2004et for us with the 3-m Shane
reflector at Lick Observatory, and Ryan Foley (University of California,
Berkeley) for reducing these spectra. The work of A.V.F.'s group at
U. C. Berkeley is supported by National Science Foundation grant AST-0307894.
Additional funding is provided by NASA through grant GO-9953 from the Space
Telescope Science Institute, which is operated by the Association of
Universities for Research in Astronomy, Inc., under NASA contract
NAS~5-26555. KAIT was made possible by generous donations from Sun
Microsystems, Inc., the Hewlett-Packard Company, AutoScope Corporation, Lick
Observatory, the National Science Foundation, the University of California, and
the Sylvia \& Jim Katzman Foundation.

\newpage

\renewcommand{\baselinestretch}{1.0}

\newpage

\begin{figure}
{\plotfiddle{sn04et.f01.new.ps}{6.4in}{0}{80}{80}{-20}{-70}}
\caption{ A $2\arcmin\times2\arcmin$ section of the KAIT $I$-band image taken on Sep. 30 of SN 2004et in
NGC 6946.  The local photometric comparison stars in the 
field of SN 2004et are labelled (see Table 1). 
North is up and east is to the left.}
\label{1}
\end{figure}

\begin{figure}
{\plotfiddle{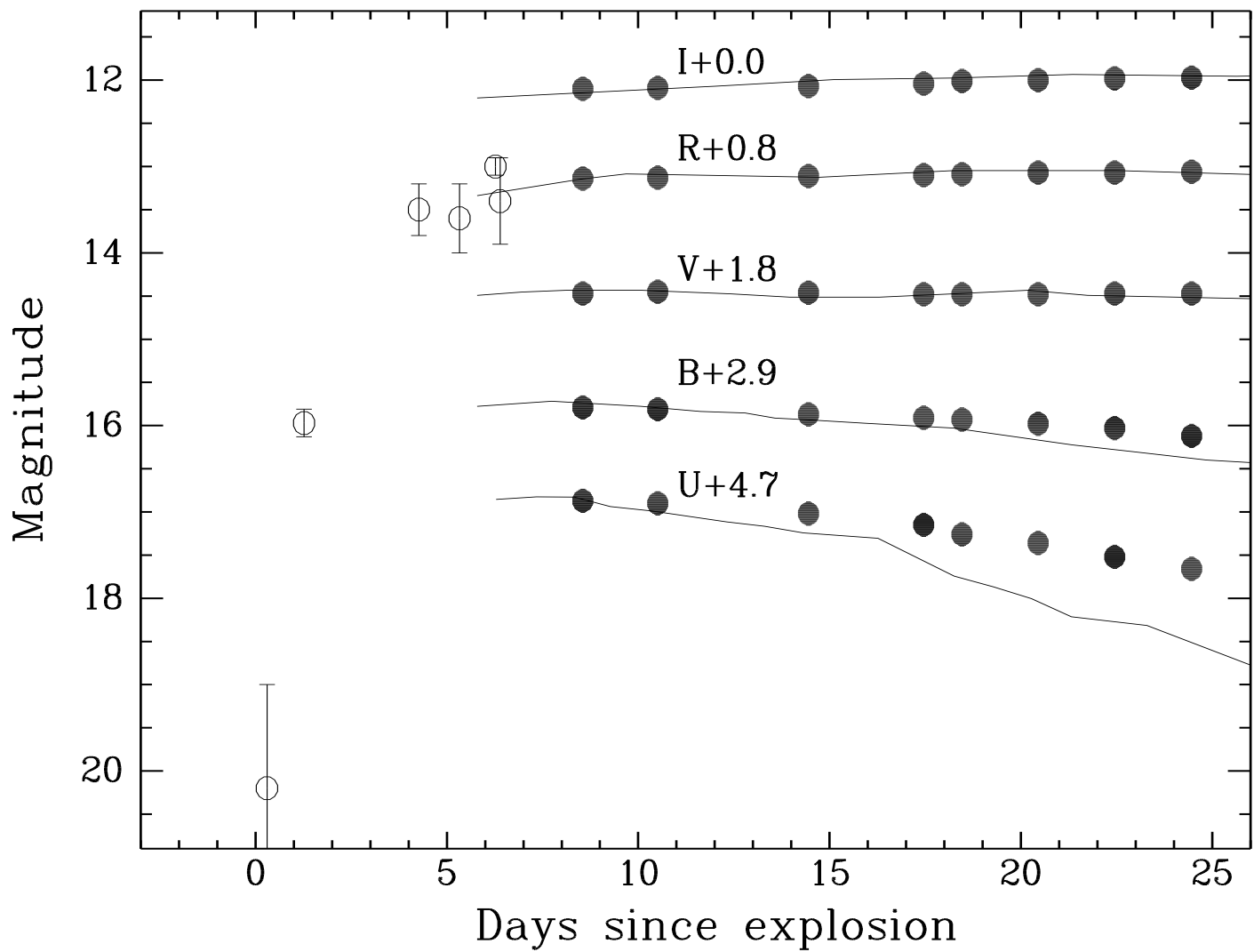}{6.4in}{-90}{80}{80}{-70}{510}}
\caption{The $UBVRI$ light curves of SN 2004et. The curves have
been shifted by the amount indicated. Also shown are light 
curves of SN 1999em ({\it solid lines}), shifted arbitrarily to
match these of SN 2004et. The pre-discovery $R$-band magnitudes of
SN 2004et reported by Klotz et al. (2004) are plotted
as {\it open circles}.}
\label{2}
\end{figure}

\begin{figure}
{\plotfiddle{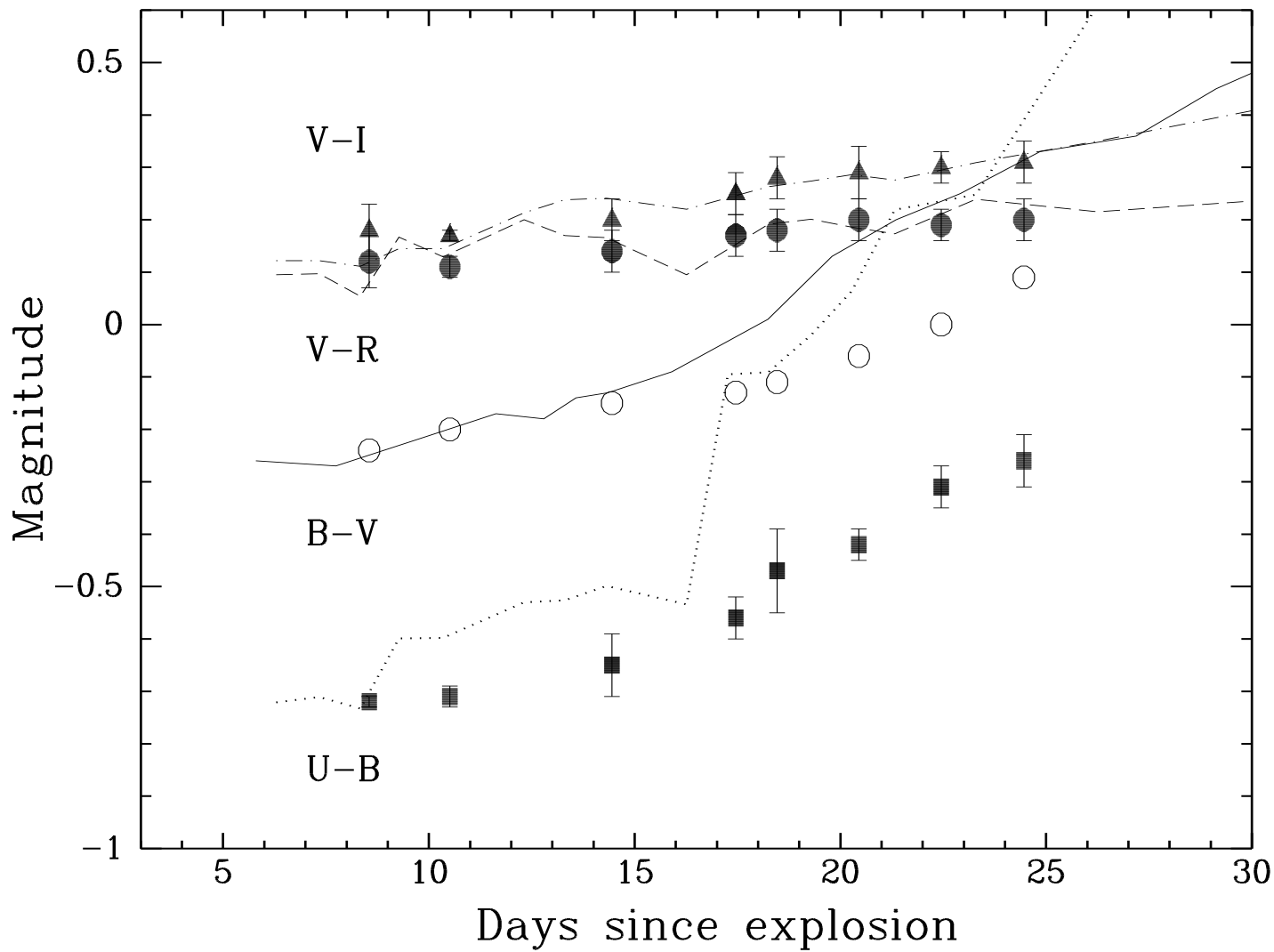}{6.4in}{-90}{80}{80}{-70}{510}}
\caption{The color curves of SN 2004et ({\it symbols}) are visually matched to 
those of SN 1999em ({\it lines}). The colors of SN 1999em have been dereddened 
by $E(B - V)$ = 0.10 mag, while the curves of SN 2004et are shifted
by 0.0 mag, $-$0.46 mag, $-$0.35 mag, and $-$0.28 mag, for the $U-B$,
$B-V$, $V-R$, and $V-I$ colors, respectively. See text for more 
details.}
\label{3}
\end{figure}

\begin{figure}
{\plotfiddle{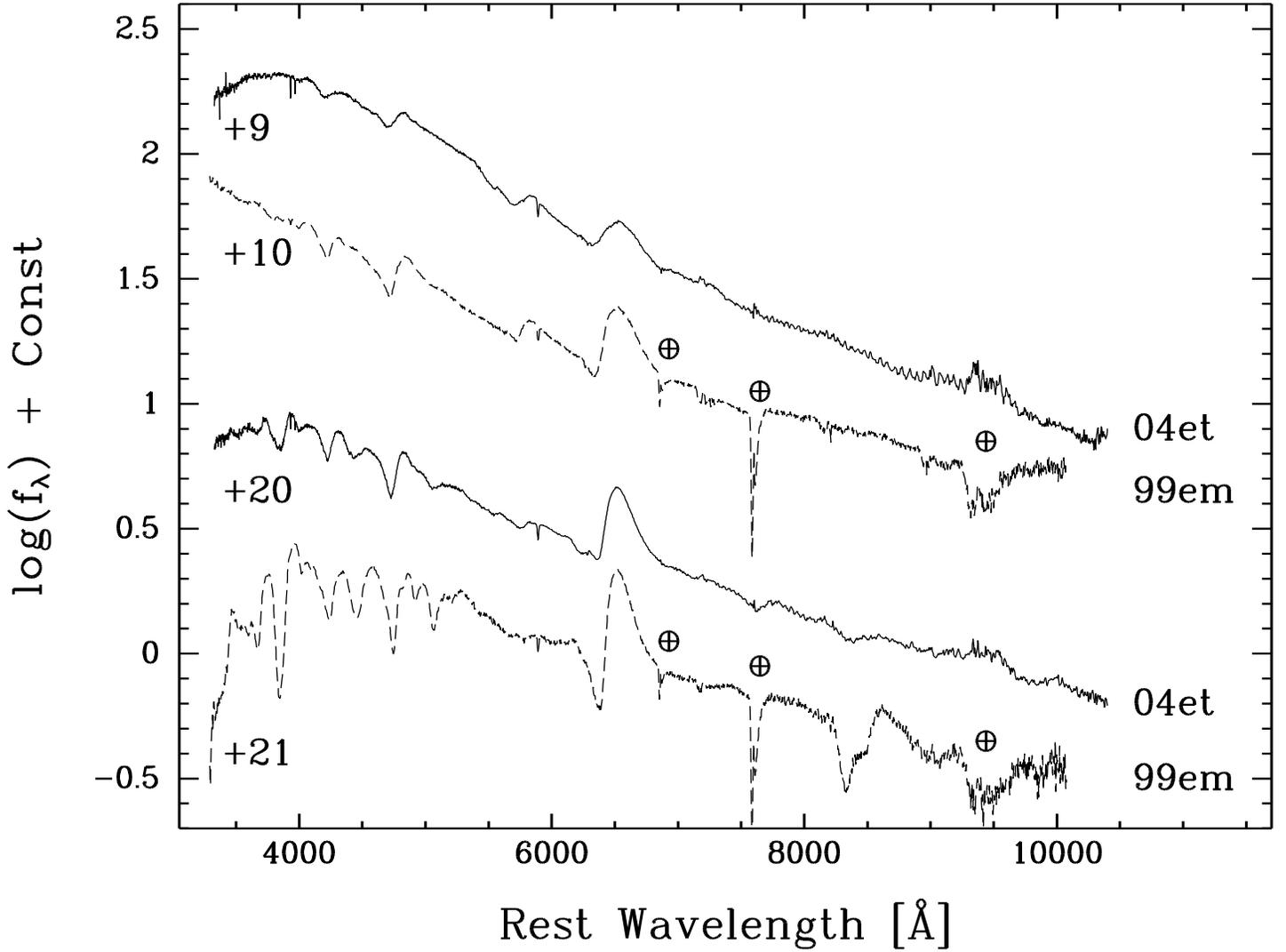}{6.4in}{-90}{90}{90}{-100}{560}}
\caption{Comparison of the SN 2004et spectra ({\it solid lines}) to
SN 1999em ({\it dotted lines}) at similar epochs. The spectra have
been corrected for reddened and host-galaxy redshift 
(see text for details). The telluric lines in the SN 1999em 
spectra are marked.}
\label{4}
\end{figure}

\begin{figure}
{\plotfiddle{sn04et.f05.new.ps}{6.4in}{0}{100}{100}{-80}{-260}}
\caption{{\it Left}:  The pre-SN CFHT $R$-band image of NGC 6946.  
This image has been carefully registered to the KAIT $I$-band 
image of the SN shown in the central panel.  North is up and
east is to the left. {\it Right}:  The SN environment in the 
CFHT image shown in greater detail. To guide the eye, the 2$\arcsec$ radius 
circle is centered on the SN position mapped from the KAIT image, although
the uncertainty in the measured SN position is much smaller ($0{\farcs}2$).
An apparent stellar object is seen near the exact center of the circle. 
The bright star used to measure relative offsets is marked by a square
(see text for details).}
\label{5}
\end{figure}

\newpage

\begin{figure}
{\plotfiddle{sn04et.f06.new.ps}{6.4in}{0}{100}{100}{-60}{-260}}
\caption{16$\arcsec \times$ 16$\arcsec$ sections, showing the
SN 2004et environment, in the CFHT $BVR{u\arcmin}{g\arcmin}{r\arcmin}$ 
images. North is up and east is to the left. The position of the 
progenitor is marked in each image.}
\label{6}
\end{figure}

\begin{figure}
{\plotfiddle{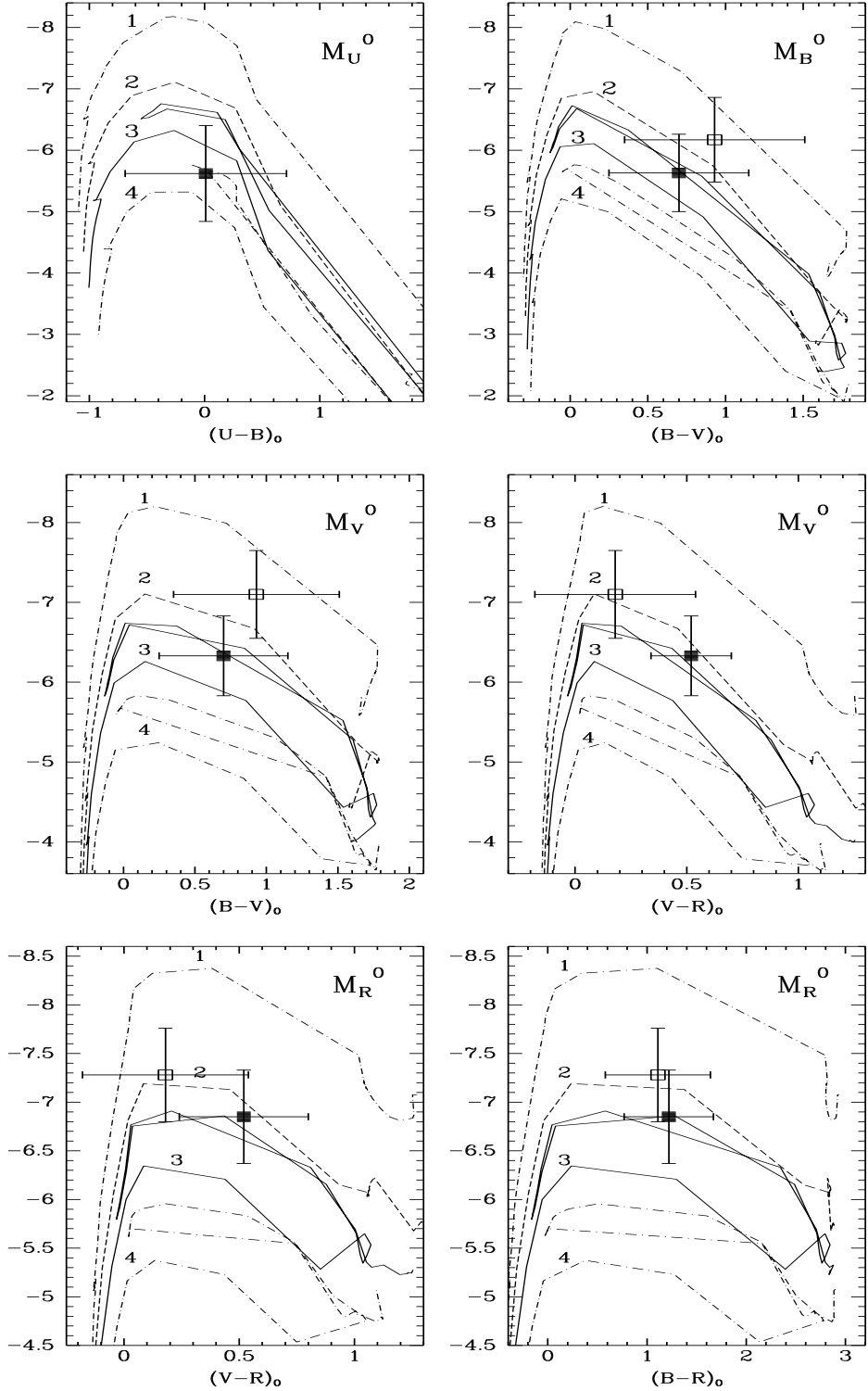}{7.4in}{0}{80}{80}{-10}{-40}}
\caption{The color-magnitude diagrams for the progenitor of
SN 2004et. The label for the absolute magnitude along the ordinate
is inside each 
panel. The {\it open square\/} represents the data derived from the 
$BVR$ images, and the {\it solid square\/} represents the Sloan
$u\arcmin g\arcmin r\arcmin$ image data. Also shown are model
stellar evolution tracks for a range of masses from Lejeune \& 
Schaerer (2001), with enhanced mass loss for the most massive stars and 
solar metallicity ($Z = 0.02$). The tracks labeled 1 to 4 
are for $M_{\rm ZAMS} = 20\ M_\odot$, $15\ M_\odot$, $12\ M_\odot$,
and $9\ M_\odot$, respectively. } 
\label{7}
\end{figure}

\begin{figure}
{\plotfiddle{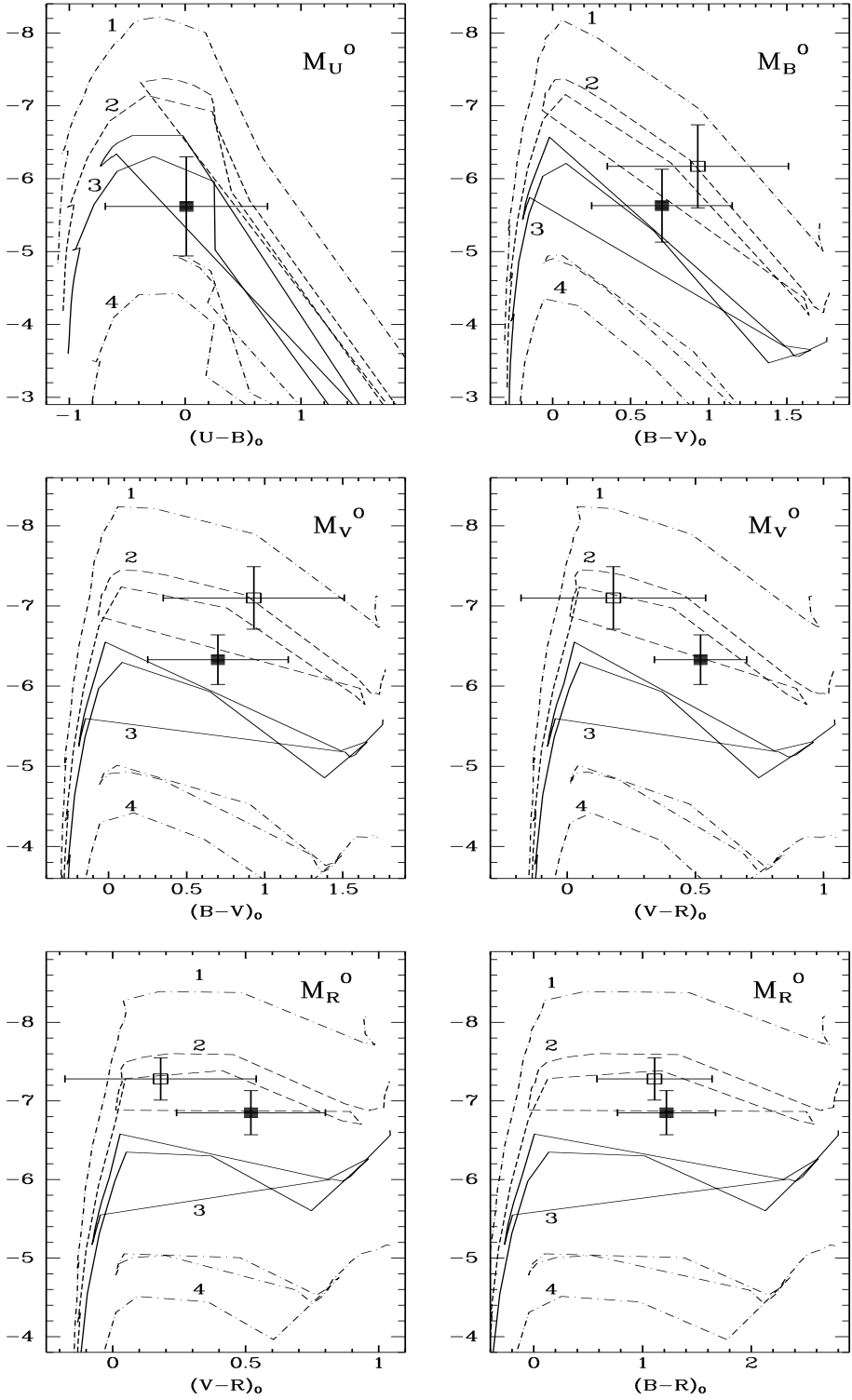}{7.4in}{0}{80}{80}{-10}{-40}}
\caption{The same as Figure 7, but for a subsolar metallicity, 
$Z = 0.008$.
The tracks labeled 1 to 4 are for $M_{\rm ZAMS} = 20\ M_\odot$,
$15\ M_\odot$, $12\ M_\odot$, and $7\ M_\odot$, respectively.}
\label{8}
\end{figure}

\newpage

\renewcommand{\arraystretch}{0.75}

\begin{deluxetable}{ccccccccc}
\ptlandscape
\tablecaption{Photometry of the Comparison Stars for SN 2004et}
\tablehead{
\colhead{ID} &\colhead{$U$} & \colhead{$B$} & \colhead{ $V$ } &
\colhead{$R$} & \colhead{$I$}&\colhead{$u\arcmin$} &\colhead{$g\arcmin$}  
 &\colhead{$r\arcmin$} 
}
\startdata
1 &\nodata & 18.31(03)& 17.09(02)& 16.41(02)&\nodata&\nodata & 17.66(04)& 
16.67(03)\\
2 &\nodata & 19.44(04)& 18.29(03)& 17.60(02)&\nodata&\nodata & 18.82(05)& 
17.90(04)\\
3 &\nodata & 19.30(03)& 18.45(03)& 18.00(03)&\nodata&\nodata & 18.82(04)& 
18.19(04)\\
4 &\nodata & 18.40(02)& 17.65(02)& 17.22(02)&\nodata&\nodata & 17.97(03)& 
17.43(03)\\
5 &\nodata & 18.31(02)& 17.37(02)& 16.87(01)&\nodata&\nodata & 17.79(03)& 
17.07(03)\\
6 &\nodata & 18.26(02)& 17.46(02)& 17.00(02)&\nodata&\nodata & 17.80(03)& 
17.21(03)\\
7 &16.62(03)&15.49(02)& 14.18(02)& 13.44(01)&12.71(02)&17.38(04)&14.79(03)& 
13.72(03)\\
8 &16.54(03)&15.54(02)& 14.33(02)& 13.62(01)&12.93(03)&17.32(04)&14.90(03)& 
13.92(03)\\
\enddata    
\tablenotetext{}{Note: Uncertainties in the last two digits 
are indicated in parentheses.}
\end{deluxetable}

\clearpage
\begin{deluxetable}{cccccc}
\tablecaption{Photometry of SN 2004et}
\tablehead{
\colhead{JD $-$ 2450000} &\colhead{$U$} & \colhead{$B$} & \colhead{ $V$ } &
\colhead{$R$} & \colhead{$I$}  
}
\startdata
3278.75& 12.17(01)&12.89(01)&12.67(03)&12.34(04)&12.10(04) \\
3280.71& 12.20(02)&12.91(01)&12.65(01)&12.33(02)&12.09(01) \\
3284.65& 12.32(06)&12.97(01)&12.66(03)&12.31(02)&12.07(03) \\
3287.66& 12.45(03)&13.01(02)&12.68(03)&12.30(02)&12.04(03) \\
3288.66& 12.56(08)&13.03(01)&12.68(04)&12.29(02)&12.01(02) \\
3290.65& 12.66(02)&13.08(02)&12.68(04)&12.27(02)&12.00(03) \\
3292.65& 12.82(04)&13.13(01)&12.67(02)&12.27(02)&11.98(02) \\
3294.66& 12.96(05)&13.22(01)&12.67(03)&12.26(02)&11.97(02) \\
\enddata    
\tablenotetext{}{Note: Uncertainties in the last two digits 
are indicated in parentheses.}
\end{deluxetable}

\begin{deluxetable}{llllccl}
\tablecaption{Journal of spectroscopic observations of SN 2004et.}
\tablehead{
\colhead{UT Date} & \colhead{$t$\tablenotemark{a}} &\colhead{Telescope}
&\colhead{Range~(\AA)\tablenotemark{b}}& \colhead{Air.\tablenotemark{c}} &
\colhead{Slit} &\colhead{Exp. (s)}  
}
\startdata
 2004-10-01 &$+$9&Lick 3-m&3300--10500 &1.1&2$\arcsec$.0 & 240 \\
 2004-10-12 &$+$20&Lick 3-m&3300--10500 &1.1&2$\arcsec$.0 & 240 \\
\enddata
\tablenotetext{a}{Days since explosion (assumed to be Sep 22.0 UT, 
JD 2453270.5), rounded to  the nearest day.}
\tablenotetext{b}{Observed wavelength range of spectrum. }
\tablenotetext{c}{Average airmass of observations.}
\end{deluxetable}

\begin{deluxetable}{lllll}
\tablecaption{Summary of CFHT observations}
\tablehead{
\colhead{Date (UT)}&\colhead{Filter}&
\colhead{Exp. (s)}&\colhead{Pixel Scale}&
\colhead{Seeing}
}
\startdata
2002 Aug. 6 & $B$ & 90$\times$5 & $0{\farcs}206$  & $0{\farcs}8$  \\
2002 Aug. 6 & $V$ & 60$\times$5 & $0{\farcs}206$ & $0{\farcs}8$  \\
2002 Aug. 6 & $R$ & 60$\times$5  & $0{\farcs}206$ & $0{\farcs}8$ \\
2003 Oct. 23 & $u\arcmin$ & 90$\times$5 & $0{\farcs}185$ & $0{\farcs}7$ \\
2003 Oct. 23 & $g\arcmin$ & 60$\times$5 & $0{\farcs}185$ & $0{\farcs}8$ \\
2003 Oct. 23 & $r\arcmin$ & 30$\times$5 & $0{\farcs}185$ & $0{\farcs}6$ \\
\enddata    
\end{deluxetable}

\begin{deluxetable}{lcccccccc}
\tablecaption{Photometry of the SN 2004et Progenitor\tablenotemark{a}}
\tablehead{
\colhead{Dataset}&\colhead{$U$} &\colhead{$\sigma(U)$} &
\colhead{$B$} &\colhead{$\sigma(B)$}  &
\colhead{$V$} &\colhead{$\sigma(V)$}  &
\colhead{$R$} &\colhead{$\sigma(R)$}  
}
\startdata
2002 Aug. &\nodata&\nodata&  24.30&0.48&  22.93&0.31&   22.50&0.18\\
2003 Oct.\tablenotemark{b}&25.12&0.58&  24.84&0.39&  23.70&0.20&   22.93&0.20\\
\hline
& & & & & & & & \\
   & $u\arcmin$ & $\sigma(u\arcmin)$  & $g\arcmin$ & $\sigma(g\arcmin)$ & & 
&$r\arcmin$ & 
$\sigma(r\arcmin)$\\
\hline
2003 Oct. &25.97&0.57&  24.18&0.21&       &    &   23.10&0.16\\
\enddata
\tablenotetext{a}{Measured from the pre-SN CFHT images.  See text for more 
details.}
\tablenotetext{b}{The $u\arcmin g\arcmin r\arcmin$ magnitudes are converted to 
$UBVR$ magnitudes.  See text for discussion of this transformation.}
\end{deluxetable}

\begin{deluxetable}{ccccc}
\tablecaption{Absolute Magnitude and Intrinsic Color of the SN 2004et 
Progenitor}
\tablehead{
\colhead{Dataset} & \colhead{$M^0_U$} & \colhead{$M^0_B$} & \colhead{$M^0_V$ } &
\colhead{$M^0_R$}
}
\startdata
2002 Aug. & \nodata&$-6.17\pm0.69$&$-7.10\pm0.55$&$-7.28\pm0.48$ \\
2003 Oct. & $-5.62\pm0.78$& $-5.63\pm0.63$&$-6.33\pm0.50$& $-6.85\pm0.48$\\
\hline
 & & & & \\
 &$(U-B)_0$&$(B-V)_0$ &$(V-R)_0$&$(B-R)_0$ \\
\hline
2002 Aug. &\nodata &$0.93\pm0.58$& $0.18\pm0.36$ & $1.11\pm0.53$ \\
 2003 Oct. & $0.01\pm0.70$ & $0.70\pm0.45$ & $0.52\pm0.28$ & $1.22\pm0.45$
\enddata    
\tablenotetext{}{Note: Uncertainties in the last two digits 
are indicated in parentheses.}
\end{deluxetable}

\begin{deluxetable}{cccc}
\tablecaption{Mass Estimates for the SN 2004et Progenitor}
\tablehead{
\colhead{CMD} & \colhead{Z} &\colhead{$M_{\rm ZAMS}$(2002 
Dataset)\tablenotemark{a}}
&\colhead{$M_{\rm ZAMS}$(2003 Dataset)\tablenotemark{b}}
}
\startdata
$[M^0_U$, $(U-B)_0]$ & 0.020 &  \nodata &   $10+8-2$ \\
$[M^0_B$, $(B-V)_0]$ & 0.020 &  $17+6-5$ &   $12+6-3$ \\
$[M^0_V$, $(B-V)_0]$ & 0.020 &  $17+6-5$ &   $12+3-2$ \\
$[M^0_V$, $(V-R)_0]$ & 0.020 &  $16\,+\,?-4$ &   $ 13+4-2$ \\
$[M^0_R$, $(V-R)_0]$ & 0.020 &  $16\,+\,?-4$ &   $ 13+4-2$ \\
$[M^0_R$, $(B-R)_0]$ & 0.020 &  $16+2-3$ &   $ 13+2-2$ \\
\hline
Median  &  0.020 &  $16+6-4$ &    $13+4-2$\\
& & & \\
\hline
$[M^0_U$, $(U-B)_0]$ & 0.008 &  \nodata &   $12+4-5$ \\
$[M^0_B$, $(B-V)_0]$ & 0.008 &  $16+6-4$ &   $ 14+4-4$ \\
$[M^0_V$, $(B-V)_0]$ & 0.008 &  $16+5-3$ &   $ 14+2-2$ \\
$[M^0_V$, $(V-R)_0]$ & 0.008 &  $15\,+\,?-2$ &   $ 14+2-2$ \\
$[M^0_R$, $(V-R)_0]$ & 0.008 &  $15\,+\,?-2$ &   $ 14+2-1$ \\
$[M^0_R$, $(B-R)_0]$ & 0.008 &  $15+2-2$ &   $ 14+2-1$ \\
\hline
Median &  0.008 &  $15+6-2$ &    $14+2-2$ \\
\enddata
\tablenotetext{a}{The $M_{\rm ZAMS}$ estimate from the 2002 image dataset 
($BVR$ images). The mass is expressed as $M_1 + M_2 - M_3$, where
$M_1$ is the estimated $M_{\rm ZAMS}$, $M_2$ is the upper limit, and $M_3$
is the lower limit.}
\tablenotetext{b}{The $M_{\rm ZAMS}$ estimate from the 2003 image dataset 
($u\arcmin g\arcmin r\arcmin$ images). }
\end{deluxetable}

\end{document}